\documentclass[final,5p,times,preprint]{elsarticle}
\usepackage[utf8]{inputenc}
\usepackage{amsfonts}
\usepackage{amsmath}
\usepackage{amssymb}
\usepackage{bm}
\usepackage{comment}
\usepackage{xcolor}
\usepackage{booktabs}
\usepackage{multirow}
\usepackage[normalem]{ulem}
\usepackage{url}
\usepackage{mathrsfs}

\journal{Physics Letters B}

\biboptions{comma,sort&compress}
\begin{document}

\begin{frontmatter}

\title{A Complex Scaling Method for Efficient and Accurate Scattering Emulation in Nuclear Reactions}



\author[Tongji]{Junzhe Liu}
\author[Tongji]{Jin Lei\corref{mail}}
\cortext[mail]{Corresponding author}
\ead{jinl@tongji.edu.cn}
\author[Tongji]{Zhongzhou Ren}

\address[Tongji]{School of Physics Science and Engineering, Tongji University, Shanghai 200092, China.}

\begin{abstract}
We present a novel scattering emulator utilizing the complex scaling method to enhance nuclear reaction analysis. This approach leverages a single set of reduced bases, allowing for efficient and simultaneous emulation across multiple channels and potential parameters, significantly reducing computational storage and accelerating calculations. Demonstrated through \(n\)+\(^{40}\)Ca and \(^{11}\)Be+\(^{64}\)Zn elastic scattering, our method achieves high accuracy and efficiency. This emulator exhibits stable and reliable performance without anomalies inherent in other techniques, showcasing its robustness. 
\end{abstract}


\end{frontmatter}

\section{Introduction}\label{sec:intro}
Nuclear reactions play a crucial role in helping us understand how nuclei are formed and the properties of dense nuclear matter, like that found in neutron stars~\cite{FRIB}. These reactions are fundamental to many astrophysical events, especially those happening in stars and during explosive occurrences like supernovae. This makes these reactions very important for astrophysics.

However, the models we use to understand these reactions depend heavily on certain assumptions, like the optical model potential, which is a way to describe how particles interact. These models need parameters that are adjusted to fit the observed data, such as elastic scattering cross sections. This fitting process introduces many uncertainties into our analysis of nuclear reactions. To deal with these uncertainties, it is vital to use robust methods for error quantification, such as Bayesian analysis, to make our results more reliable and understandable~\cite{PhysRevLett.122.232502,ROSE}.

In the realm of Bayesian analysis, the necessity to perform calculations repeatedly across randomly sampled parameter sets poses significant computational demands. To mitigate this, researchers have developed emulators—computationally efficient algorithms designed to approximate precise calculations with high accuracy. Initially introduced for solving eigenvalue problems in bound states~\cite{EC_original,KONIG2020135814}, the concept quickly expanded to address scattering problems~\cite{FURNSTAHL2020135719}. Emulators have proven particularly useful in nuclear physics, where they can significantly speed up the analysis process without sacrificing accuracy.

Several types of nuclear scattering emulators have been developed. For example, emulators based on the Kohn variational principle~\cite{FURNSTAHL2020135719,PhysRevC.105.064004}, other variational methods~\cite{DRISCHLER2021136777,MELENDEZ2021136608}, and the $R$-matrix theory~\cite{PhysRevC.103.014612,PhysRevC.106.024611} have shown impressive accuracy and efficiency. These emulators can quickly provide results that are very close to those obtained from detailed and time-consuming calculations.

However, there are some limitations. One major issue is that most emulators are designed to work with specific angular momentum states, which means they cannot easily be used for different angular momentum settings. This specialization increases the amount of storage needed and the computational time required, as the unique properties of each angular momentum state are not fully utilized. This limitation reduces the overall utility of these emulators, highlighting the need for more flexible and general approaches.

In this article, we introduce a novel scattering emulator that leverages the complex scaling method to address the limitations of previous emulators~\cite{kruppaprc,coloss,CARBONELL201455}. The complex scaling operation transforms the oscillatory boundary conditions of the scattering wave function into exponentially decaying ones~\cite{kruppaprc,PhysRevC.84.034002}. This transformation helps eliminate discrepancies in boundary conditions across channels with different angular momentum.

Our new approach utilizes a single set of reduced basis generated through the eigenvector continuation method~\cite{EC_original}, allowing for simultaneous emulation across different channels and potential parameters. This significantly reduces the required computational storage. Additionally, all channels use the same Hamiltonian matrix relative to this basis, differing only in the centrifugal barrier term. This uniformity greatly enhances computational efficiency.

To test the effectiveness of our emulator, we apply it to elastic scattering scenarios involving $ n + ^{40}\text{Ca} $ and $ ^{11}\text{Be} + ^{64}\text{Zn} $, and we quantify the emulator's relative errors. Notably, our emulator avoids the singularity issues that are often seen in some other variational methods, highlighting the robustness and reliability of our approach.

The paper is organized as follows: In Sec.~\ref{sec:theory}, we provide a brief overview of the complex scaling method and the development of our scattering emulator. Sec.~\ref{sec:results} presents some applications of our emulator, and an anomaly detection test. Finally, we conclude with a summary and discuss future research directions in Sec.~\ref{sec:sum}.

\section{Formalism}\label{sec:theory}
The collision process of two atomic nuclei can be effectively described using the Optical Model Potential (OMP), a powerful tool in nuclear reaction theories. Typically, the OMP is assumed to have a Woods-Saxon form with specific parameters. These parameters are fine-tuned to match experimental elastic scattering cross-sections. The radial Schrödinger equation incorporating a local OMP for a given angular momentum $\ell$ is expressed as:
\begin{equation}
    \left[-\frac{\hbar^2}{2\mu} \frac{\mathrm{d}^2}{\mathrm{d}r^2} + \frac{\hbar^2}{2\mu} \frac{\ell(\ell+1)}{r^2} + U_N(r;\boldsymbol{\omega}) \right] \psi_\ell(r) = E \psi_\ell(r),
    \label{eq:original_schrodinger}
\end{equation}
where $\mu$ is the reduced mass of the system, $E$ is the kinetic energy in the center-of-mass frame, and $U_N$ is the optical potential which depends on a parameter configuration denoted by $\boldsymbol{\omega}$. The optical model potential in this study is defined as:
\begin{equation}
\begin{aligned}
    U_N(r;\boldsymbol{\omega}) = & V_C(r;R_c) + V_0 \mathscr{Y}\left(r; R_R, a_R\right) \\
    & + i W_0 \mathscr{Y}\left(r; R_W, a_W\right) + i W_S \frac{\mathrm{d}}{\mathrm{d}r} \mathscr{Y}\left(r; R_{ws}, a_{ws}\right),
\end{aligned}
\end{equation}
where $\mathscr{Y}(r;R,a)\equiv (1+\mathrm{exp}((r-R)/a))^{-1}$ is the Woods-Saxon (WS) function, $V_C$ is the Coulomb interaction, the second and third terms are the real and imaginary volume terms, and the last term is the imaginary surface term. The parameter configuration is:
\begin{equation}
    \boldsymbol{\omega} = \{V_0, R_R, a_R, W_0, R_W, a_W, W_S, R_{ws}, a_{ws}, R_c\},
\end{equation}
which characterizes the depth, radius, and diffuseness of the real and imaginary volumes, imaginary surface terms, and the charge radius for the Coulomb interaction. In this study, we treat the angular momentum quantum number $\ell$ as a parameter and define the parameter-dependent Hamiltonian as:
\begin{equation}
    H(r,\boldsymbol{\Omega}) = -\frac{\hbar^2}{2\mu} \frac{\mathrm{d}^2}{\mathrm{d}r^2} + \frac{\hbar^2}{2\mu} \frac{\ell(\ell+1)}{r^2} + U_N(r;\boldsymbol{\omega}),
\end{equation}
where $\boldsymbol{\Omega} = \{\boldsymbol{\omega}, \ell\}$ is the total parameter configuration we consider in this work. It should be noted that in some conventional methods for solving Eq.~(\ref{eq:original_schrodinger}), the angular momentum quantum number $\ell$ serves not only as a parameter in the Hamiltonian but also specifies the boundary condition of the scattering wave function. This presents significant challenges when dealing with states of different angular momentum. To remove the dependence on different boundary conditions and treat $\ell$ just as a parameter, we apply the complex scaling method (CS) to solve Eq.~(\ref{eq:original_schrodinger}). 
In the complex scaling method, the wave function is first separated into two parts
\begin{equation}
    \psi_{\ell}(r) = e^{i \sigma_\ell} F_\ell(\eta, kr) + \psi_{\ell}^{\text{sc}}(r),
\end{equation}
where $\sigma_\ell$ is the Coulomb phase shift, $F_\ell$ is the regular Coulomb function, $\eta$ is the Sommerfeld parameter, $k = \sqrt{2\mu E}/\hbar$ is the wave number, and $\psi_{\ell}^{\text{sc}}(r)$ is referred to as the scattered part of the wave function.
The coordinate of $\psi_{\ell}^{\text{sc}}(r)$ is then rotated by an angle $\theta$ to get the complex-scaled scattered part of the wave function 
\begin{equation}
    \psi_{\ell}^{\text{sc},\theta}(r)=e^{i\theta/2}\psi_{\ell}^{\text{sc}}(re^{i\theta}),
\end{equation} 
and it satisfies the following inhomogeneous equation:
\begin{equation}
    \left[E - H^{\theta}(r;\boldsymbol{\Omega})\right] \psi_{\ell}^{\text{sc},\theta}(r) = e^{i\theta/2} e^{i \sigma_\ell} \tilde{U}_N(re^{i\theta};\boldsymbol{\Omega}) F_\ell(\eta, kre^{i\theta}),
    \label{eq:cs_driven}
\end{equation}
where $H^{\theta}$ is the complex-scaled Hamiltonian, and $\tilde{U}_N$ is the modified interaction defined as:
\begin{equation}
    \tilde{U}_N(re^{i\theta};\boldsymbol{\Omega}) = U_N(re^{i\theta};\boldsymbol{\Omega}) - \frac{z_1 z_2 e^2}{re^{i\theta}},
\end{equation}
where $z_1$ and $z_2$ are the charge numbers of the two nuclei. 
Upon separating the wave function, the partial wave expanded nuclear scattering amplitude can be decomposed into two terms as well:
\begin{equation}
    f_\ell = f_\ell^{\text{Born}} + f_\ell^{\text{sc}},
\end{equation}
where the Born term of the scattering amplitude is defined as:
\begin{equation}
f_\ell^{\text{Born}}
= -\frac{1}{E} \int dr 
F_\ell(\eta, kr) \tilde{U}_N(r) F_\ell(\eta, kr),
\label{eq:fborn}
\end{equation}
and the correction to the Born term of the scattering amplitude can be derived with the complex-scaled scattered part of the wave function using Cauchy theorem: 
\begin{equation}
f_\ell^{\text{sc}}(k) 
= 
-\frac{e^{i \theta / 2} e^{-i \sigma_\ell}}{E}  \int_0^{\infty} dr F_\ell\left(k r e^{i \theta}\right) 
\tilde{U}_N\left(re^{i\theta};\boldsymbol{\Omega}\right)
\psi_{\ell}^{\text{sc}, \theta}(r).
\label{scamp_cs}
\end{equation}
The total nuclear scattering amplitude, denoted as $\mathscr{F}_N$, can be obtained as a sum over the partial wave expanded amplitudes:
\begin{equation}
    \mathscr{F}_N(\cos{\theta}) = \sum_\ell (2\ell+1) e^{2i\sigma_\ell} f_\ell P_\ell(\cos{\theta}),
\end{equation}
where $P_\ell$ is the Legendre polynomial.
The angle distribution of the cross section is given by the modular square of the total scattering amplitude:
\begin{equation}
    \frac{d\sigma}{d\Omega} = \left|\mathscr{F}_N + \mathscr{F}_{\mathrm{C}}\right|^2,
    \label{eq:diffxsection}
\end{equation}
where $\mathscr{F}_C$ is the Coulomb scattering amplitude.
For a more detailed derivation of the complex scaling method for solving scattering problems and the calculation of the cross section, one may refer to Ref.~\cite{coloss}, in which we have developed a computer code, COLOSS, to solve the two-body scattering problem using the complex scaling method. 

Since $\psi_{\ell}^{\text{sc},\theta}$ is exponentially decaying after the complex scaling operation, we are able to use a complete set of square-integrable functions as the variational basis to expand it.
For a direct solution of the complex scaling method for a given parameter set of the Schrödinger equation, which can be called the exact solution, we expand $\psi_{\ell}^{\text{sc},\theta}$ with $x$-regularized Lagrange-Laguerre functions \cite{baye_lagrange-mesh_2015}:
\begin{equation}
    \psi_{\ell}^{\text{sc},\theta}(r) = 
    \sum_j^{N_r} c_j g_j(r),
    \label{eq:ori_expansion}
\end{equation}
where $c_j$ represents the expansion coefficient, $g_j(r)$ denotes the $x$-regularized Lagrange-Laguerre function, and $N_r$ is the number of basis functions. By substituting Eq.~(\ref{eq:ori_expansion}) into Eq.~(\ref{eq:cs_driven}) and projecting on $g_i$, the inhomogeneous ordinary differential equation can be transformed into a set of linear equations:
\begin{equation}
    \sum_j \left[E \mathbf{N}_{ij} - \mathbf{H}^{\theta}_{ij}(\boldsymbol{\Omega})\right] c_j(\boldsymbol{\Omega}) = b_i^{\theta}(\boldsymbol{\Omega}),
    \label{eq:linear_eq}
\end{equation}
where $\mathbf{N}_{ij}$ is the inner product matrix of the basis functions, $\mathbf{H}^{\theta}_{ij}$ is the matrix element of the Hamiltonian given by:
\begin{equation}
    \mathbf{H}^{\theta}_{ij} = \int \mathrm{d}r \, g_i(r) H^{\theta}(r;\boldsymbol{\Omega}) g_j(r),
    \label{eq:Hmat}
\end{equation}
and $b_i^{\theta}(\boldsymbol{\Omega})$ is the inhomogeneous term defined as:
\begin{equation}
    b_i^{\theta}(\boldsymbol{\Omega}) = \int \mathrm{d}r \, g_i(r) e^{i\theta/2} e^{i \sigma_\ell} \tilde{U}_N(re^{i\theta};\boldsymbol{\Omega}) F_\ell(\eta, kre^{i\theta}).
    \label{eq:inhomo_term}
\end{equation}
With the solution of the linear equation, the correction to the Born term of the scattering amplitude can be simplified as \cite{coloss}:
\begin{equation}
f^{\text{sc}}_\ell 
= 
-\frac{e^{-2i \sigma_\ell}}{E}  
\sum_i
c_i(\boldsymbol{\Omega})
b_i^{\theta}(\boldsymbol{\Omega}).
\label{eq:fsc_lineareq}
\end{equation}

The linear equation form of Eq.~(\ref{eq:linear_eq}) serves as a natural starting point for constructing an emulator using the eigenvector continuation method \cite{EC_original,FURNSTAHL2020135719,DRISCHLER2021136777}, known as a reduced basis method \cite{quarteroni2015reduced}. To develop our new emulator, which can span multiple channels and potential parameters, we initially conduct exact calculations for $N_s$ different parameter sets $\{\boldsymbol{\Omega}_i\}_{i=1}^{N_s}$ and obtain exact solutions $\{\mathbf{c}^{(i)}\}_{i=1}^{N_s}$. To further reduce the basis size, we perform principal component analysis (PCA) on these solutions to extract the first $n$ principal components:
\begin{equation}
    \{\mathbf{x}^{(k)}\}_{k=1}^{n} = \text{PCA}\left[ \{\mathbf{c}^{(i)}\}_{i=1}^{N_s} \right].
\end{equation}
The new reduced basis can be constructed using these principal components as follows:
\begin{equation}
    \phi_k(r) = \sum_{i=1}^{N_r} x^{(k)}_i g_i(r),
\end{equation}
where $\phi_k$ represents the $k$th reduced basis, and $x^{(k)}_i$ denotes the $i$th expansion coefficient of the $k$th principal component vector. All the Lagrange functions in Eq.~(\ref{eq:ori_expansion}), (\ref{eq:Hmat}), and (\ref{eq:inhomo_term}) can then be replaced by $\{\phi_k\}_{k=1}^{n}$ as the new basis, and the approximated linear equation, similar to Eq.~(\ref{eq:linear_eq}), to be solved for emulation can be derived. The scattering amplitude can be emulated according to Eq.~(\ref{eq:fsc_lineareq}) with the solutions of the approximated linear equation and its inhomogeneous terms. In other words, the emulator approximates the linear equation for a given target parameter set via projection onto the reduced basis, facilitating a straightforward solution. Consequently, the number of basis functions can be reduced from $N_r$ to $n$, leading to a reduced time complexity from $O(N_r^3)$ to $O(n^3)$ for solving the linear equation.

In our approach, emulators for various partial waves can utilize the same set of reduced bases, resulting in substantial savings in computer storage and accelerated computation of matrix elements. This efficiency arises from the fact that the centrifugal barrier term remains the sole distinction among matrix elements in different channels. It is crucial to emphasize that in this method, no matrix inversion is required. This characteristic not only enhances numerical stability and efficiency but also mitigates singularity issues, distinguishing it from emulators based on the Kohn variational principle \cite{FURNSTAHL2020135719,DRISCHLER2021136777}.

\section{Results}\label{sec:results}
\subsection{Application to $n$+$^{40}$Ca elastic scattering}
We apply our emulator first to $ n+^{40}\text{Ca} $ elastic scattering at $ E=20 $ MeV in the center-of-mass frame. We include partial-wave channels with angular momentum $\ell \leq 10$ to ensure convergence.

To construct the reduced bases for our emulator, we initially select 100 parameter sets, $\{\boldsymbol{\Omega}_i\}_{i=1}^{100}$, randomly generated within a $\pm20\%$ range centered on the optimal values derived from the Koning-Delaroche (KD) parameterization \cite{koning2003local}. We perform this selection using Latin hypercube sampling (LHS) \cite{LHS}, and we confine our parameter space to this range. Angular momentum quantum numbers, $\ell$, are randomly chosen as integers from 0 to 10.\footnote{Including parameter sets with low angular momentum in the training set is essential, while excluding higher $\ell$ values does not significantly reduce accuracy. This is because the centrifugal barrier term for states with high angular momentum dominates over the optical potential, making it challenging for the emulator to capture the analytic properties of the optical potential effectively.} We perform exact calculations for these training parameter sets, yielding 100 exact solutions. Subsequently, principal component analysis (PCA) is applied to these solutions to derive reduced bases of 10, 12, and 14 dimensions, which are used to train our emulator.

\begin{figure}[h]
    \centering
    \includegraphics[width=0.9\linewidth]{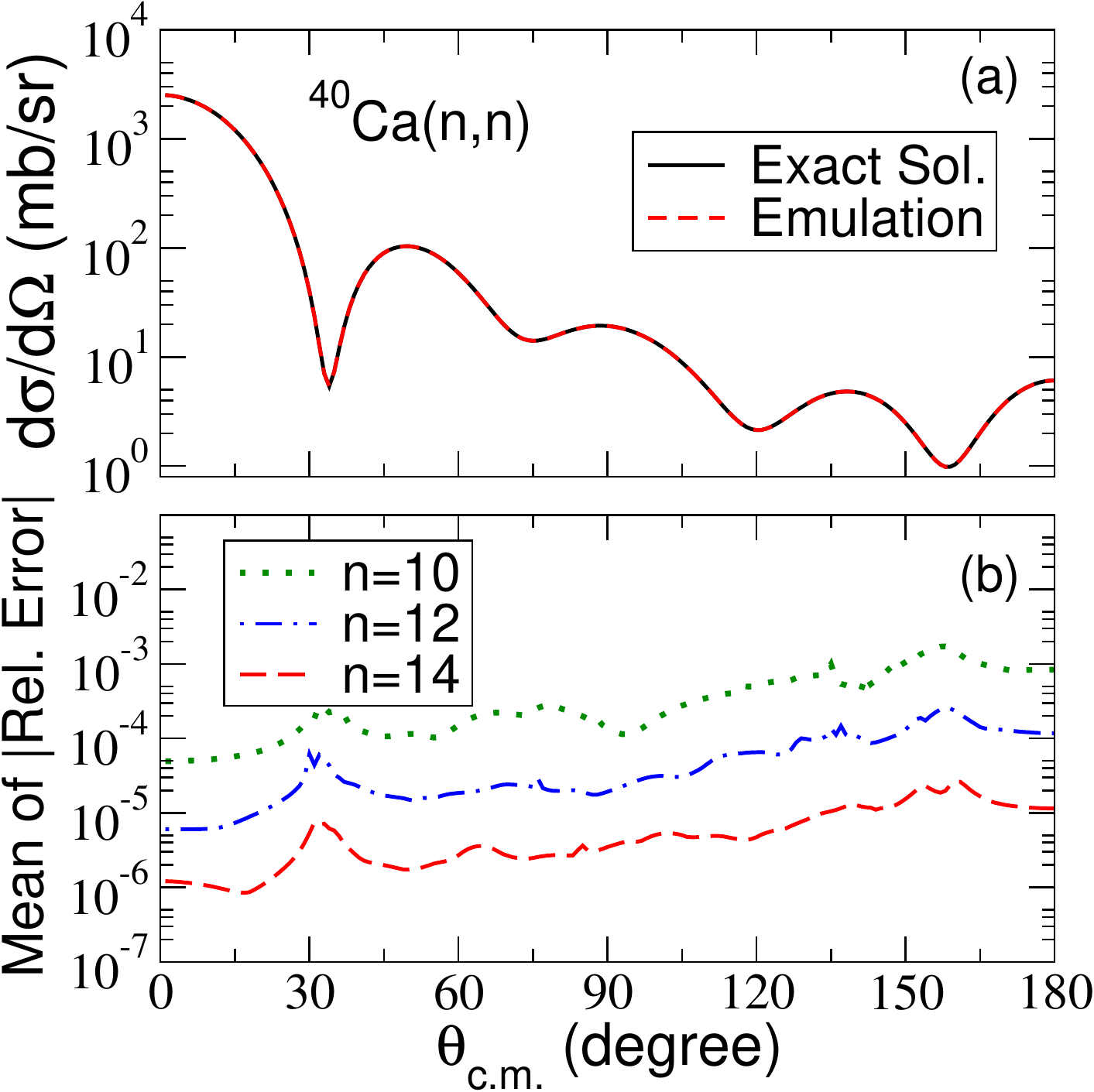}
    \caption{(a) Exact and emulated elastic scattering differential cross sections of the $n+^{40}\text{Ca}$ reaction at $E_{\text{c.m.}}=20$ MeV as a function of scattering angles. (b) Mean absolute relative errors of 100 sets of target parameters for several sizes of the reduced bases as a function of scattering angles. See text for more details.}
    \label{fig:n40Ca_accuracy_test}
\end{figure}

With these reduced bases, we first emulate the cross section on the parameters provided by the Koning-Delaroche (KD) parameterization~\cite{koning2003local} to make a comparison with the exact results. To robustly assess the numerical accuracy and stability of our method, it is crucial to conduct multiple emulations. Consequently, we generated another 100 sets of target parameters across the entire parameter space using the Latin Hypercube Sampling (LHS) method and repeated the emulation on these sets to examine the variability and reliability of the errors. The mean absolute relative error of 100 results, computed as follows, serves as a criterion for assessing our method:
\begin{equation}
\epsilon = \sum_i \frac{|\sigma_{\text{E},i} - \sigma_{0,i}|}{\sigma_{0,i}},
\label{eq:error}
\end{equation}
where $\epsilon$ represents the mean of absolute relative errors, and $\sigma_{\text{E},i}$ and $\sigma_{0,i}$ denote the emulated and high-fidelity cross-section for the $i$th target parameter set. By averaging the errors over multiple emulations, we can obtain a more reliable measure of the emulator's performance and provide a robust evaluation of numerical accuracy and stability. The results are shown in Fig.~\ref{fig:n40Ca_accuracy_test}.

In panel (a) of Fig.~\ref{fig:n40Ca_accuracy_test}, the exact and emulated differential cross sections as a function of the scattering angle are illustrated by red dashed and black solid lines. Values of the parameters are taken from the Koning-Delaroche parameterization ~\cite{koning2003local}. We have adopted 14 reduced basis states for the emulation and 60 $x$-regularized Lagrange-Laguerre basis for the exact calculation. It can be observed that our emulator effectively replicates the angular distribution of the cross section, demonstrating the emulator's accuracy.

In panel (b) of Fig.~\ref{fig:n40Ca_accuracy_test}, the mean absolute relative errors for 100 sets of target parameters are plotted as a function of the scattering angle for several reduced basis sizes. The errors associated with reduced bases of 10, 12, and 14 dimensions are represented by green dotted, blue dot-dashed, and red dashed lines, respectively. The figure demonstrates that the accuracy of the emulator significantly improves—approximately by an order of magnitude—with each two-dimensional increment in the reduced basis size, underscoring the emulator's rapid convergence and high accuracy. Notably, error peaks occur at angles where the cross section is minimal and challenging to measure accurately. This highlights the emulator's capability to deliver highly precise solutions across a broad parameter space, even while operating within a considerably smaller subspace and using a consistent set of reduced bases across different angular momentum channels. With the performance testing across the entire parameter space, we can estimate the uncertainty introduced by the emulator by assuming it has a constant relative error, as detailed in Ref.~\cite{ROSE}. 

\subsection{Application to $^{11}$Be+$^{64}$Zn elastic scattering}
In a practical demonstration of our emulator's capabilities, we examine the elastic scattering of $^{11}$Be on a $^{64}$Zn target at $E_{\text{c.m.}}=24.5$ MeV, which is approximately 1.4 times the Coulomb barrier \cite{PhysRevLett.105.022701}. Extensive experimental research has been conducted on $^{11}$Be-induced reactions to elucidate the reaction mechanisms of halo nuclei \cite{PhysRevLett.105.022701, PhysRevLett.118.152502, PhysRevC.85.054607, PhysRevC.105.034602}. 

To achieve a converged result, calculations require up to $\ell_{\text{max}}=60$. Traditional emulator methods necessitate preparing 61 distinct sets of reduced bases and repeating the computation of Hamiltonian matrix elements 61 times. 

Our emulator offers a significant advantage by using a single set of reduced bases to simulate 61 channels with varying angular momentum and potential parameters simultaneously. This approach reduces the storage requirements for different channel bases by approximately an order of magnitude compared to previous studies. Additionally, nearly identical matrix elements are used across all channels (except for the centrifugal barrier terms), which significantly expedites the computational process.

To construct the reduced basis for the emulator, we initially selected 200 sets of training parameters, denoted as $\{\boldsymbol{\Omega}_i\}_{i=1}^{200}$. These potential parameters were randomly generated using Latin Hypercube Sampling (LHS) within a $\pm20\%$ interval of the best-fit values reported in Ref.~\cite{PhysRevLett.105.022701}, with the parameter space confined to this range. The angular momentum quantum numbers, $\ell$, were chosen as random integers between 0 and 60. Principal Component Analysis (PCA) was subsequently applied to the solutions of these parameter sets to derive 20, 25, and 30 reduced bases for training the emulator.

\begin{figure}[h]
    \centering
    \includegraphics[width=0.9\linewidth]{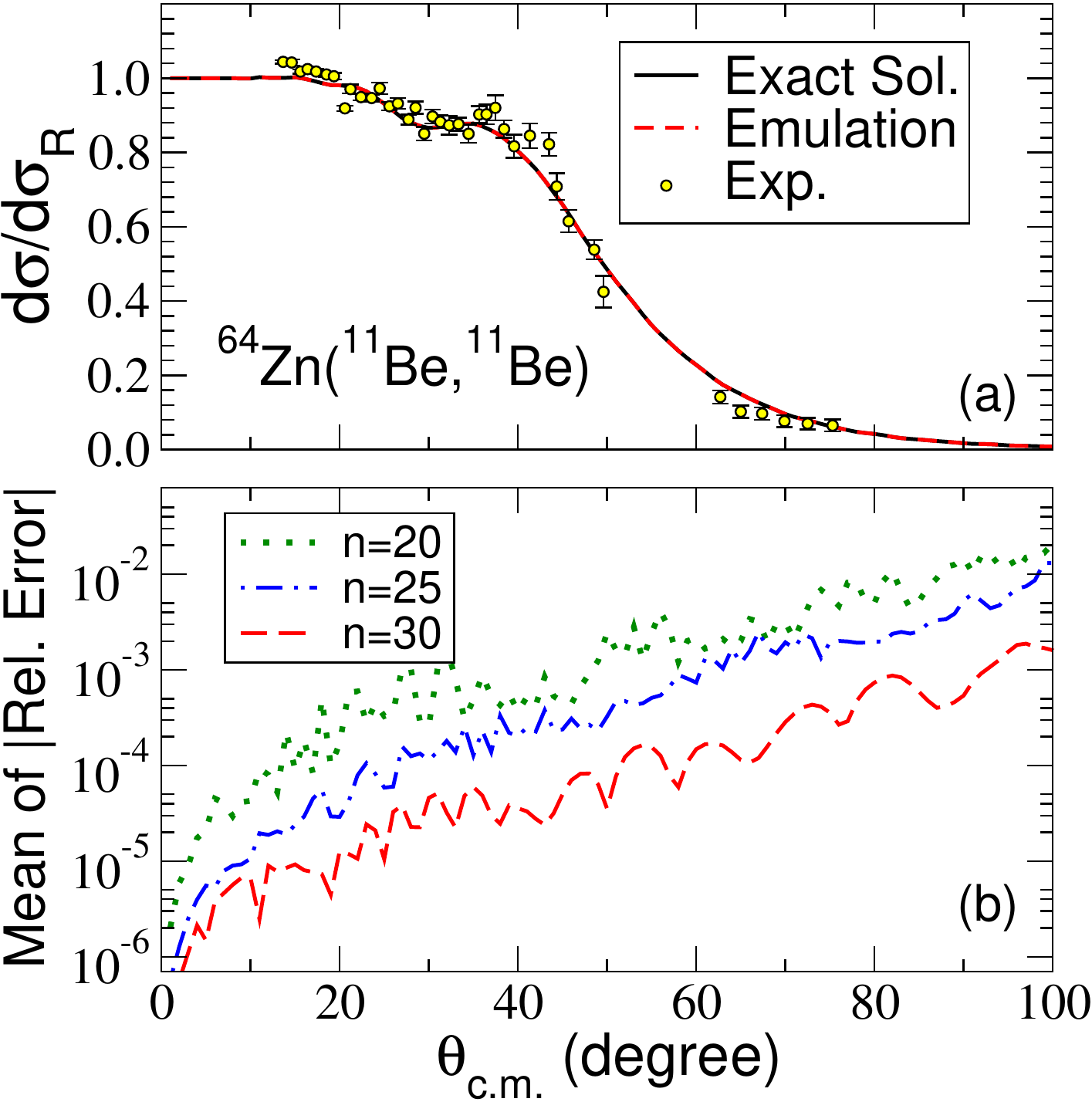}
    \caption{(a) Exact and emulated elastic scattering cross section of $^{11}$Be+$^{64}$Zn reaction at $E_{\text{c.m.}}=24.5$ MeV. The data taken from Ref.~\cite{PhysRevLett.105.022701} is also included. (b) Mean absolute relative errors of 100 sets of target parameters with respect to the exact results for several sizes of reduced basis. See text for more details.}
    \label{fig:11Be64Zn_accuracy_test}
\end{figure}

We initially applied the emulator to the best-fit parameters from Ref.~\cite{PhysRevLett.105.022701} to facilitate a direct comparison with both the exact results and the experimental data. Following this initial test, the emulator's performance was evaluated on 100 sets of target parameters, also randomly generated using LHS across the entire parameter space. The mean absolute relative errors with respect to the exact solutions, as specified in Eq.~(\ref{eq:error}), were then calculated. The results of these tests are presented in Fig.~\ref{fig:11Be64Zn_accuracy_test}.

Figure~\ref{fig:11Be64Zn_accuracy_test} (a) displays the differential cross sections as a function of the scattering angle, comparing exact, emulated, and experimental results. For this analysis, 30 reduced basis states were utilized for the emulation, while 100 $x$-regularized Lagrange-Laguerre basis were employed for the exact calculations. In this panel, the exact results (black solid line) and the emulation results (red dashed line) are indistinguishable, with both lines perfectly overlapping and accurately reproducing the experimental data (yellow circles). This outcome confirms that the complex scaling-based emulator can effectively emulate across a wide range of angular momentum values.

Figure~\ref{fig:11Be64Zn_accuracy_test} (b) displays the mean of absolute relative errors of 100 different sets of parameters for various sizes of the reduced basis. The errors are plotted as a function of the scattering angle, with different basis sizes represented by distinct line styles: 20 basis states by a green dotted line, 25 basis states by a blue dot-dashed line, and 30 basis states by a red dashed line. A clear trend of rapid convergence is evident; the error diminishes substantially as the basis size increases. Notably, at scattering angles up to approximately $70^\circ$ — where data are available — the relative errors remain well around $10^{-3}$ for the 20 reduced basis scenario, significantly surpassing the typical experimental uncertainty of 5\%. These results underscore our emulator's high efficiency and accuracy in simultaneously emulating across various channels and potential parameters across the entire parameter space with a single set of reduced bases. This demonstrates its substantial practicality for handling realistic scenarios in nuclear reactions research.

\subsection{No detection of anomaly}
Previous studies on developing emulators for scattering problems have mainly used various variational principles, such as the Kohn variational principle (KVP) and others cited in works like~\cite{kvp,schiwingervp,newton2013scattering}. The KVP, which was the first to be applied, often faces computational challenges. Specifically, the matrices involved can become ill-conditioned at larger basis sizes, leading to numerical instabilities known as the "Kohn anomaly" at certain energy values~\cite{DRISCHLER2021136777,PhysRevA.40.6879,zhang1988quantum}. These anomalies create significant difficulties, prompting various efforts to address these instabilities.

In contrast, our research using the complex scaling method shows a clear advantage: no such anomalies were observed. The new emulator we developed is remarkably stable across a range of energies and basis sizes, avoiding the issues inherent in the KVP approach. This stability enhances the reliability of the emulator and provides a robust tool for error quantification.

To enable a direct comparison, we replicate the calculation from Ref.~\cite{DRISCHLER2021136777}, focusing on the scattering state of $^{10}$Be-$n$ described by a real Woods-Saxon potential with a spin-orbit term:
\begin{equation}
V(r)=-V_0 \mathscr{Y}(r ; R, a) + \bm{\ell} \cdot \mathbf{s} \frac{V_{\mathrm{LS}}}{r} \frac{\mathrm{d}}{\mathrm{d} r} \mathscr{Y}(r ; R, a),
\end{equation}
where $ \mathscr{Y} $ is the Woods-Saxon function, and the parameter values are taken from Ref.~\cite{PhysRevC.70.064605}. This real potential is fitted to reproduce the $\mathrm{d}_{5/2}$ resonance at 1.274 MeV. 

Previous studies reported that the KVP method exhibited an anomaly in the phase shifts of the $\mathrm{d}_{5/2}$ state at around 8.4 MeV in the center-of-mass frame. To investigate whether our method encounters any similar anomalies, we calculate and emulate the phase shift of this $\mathrm{d}_{5/2}$ state across a center-of-mass energy range from 0.02 MeV to 10 MeV, using an energy step size of 0.02 MeV.
The parameter set we emulate is:
\begin{equation}
\boldsymbol{\Omega} = \{V_0,V_{\mathrm{LS}},a,R\}.
\end{equation}
We select 15 sets of training parameters, $\{\boldsymbol{\Omega}_i\}_{i=1}^{15}$, with each value randomly generated within a $\pm20\%$ interval of the best-fit values in Ref.~\cite{PhysRevLett.105.022701}, and we confine our parameter space within this range. Using exact solutions on these parameter sets, we apply principal component analysis to establish 8 reduced basis states for the emulator. 

\begin{figure}[h]
    \centering
    \includegraphics[width=0.9\linewidth]{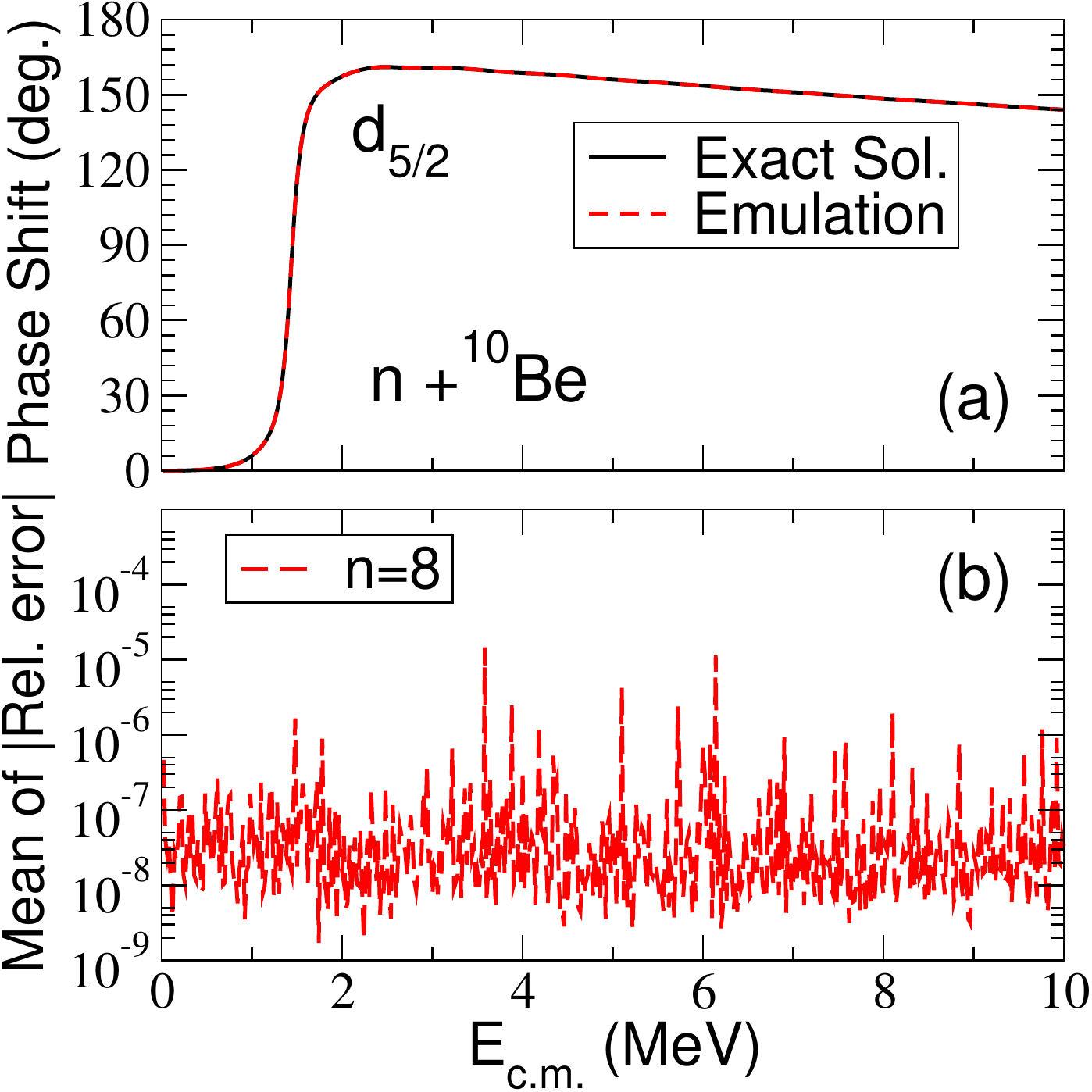}
    \caption{(a) $\mathrm{d}_{5/2}$ phase shifts of $n$+$^{10}$Be scattering as a function of the center-of-mass energy. (b) Mean of absolute relative error with respect to the exact results. See text for more details.}
    \label{fig:delta_n10Be}
\end{figure}

We initially emulate the phase shifts across the energy spectrum using the best-fit values from Ref.~\cite{PhysRevLett.105.022701}. To rigorously evaluate the emulator's accuracy, we perform the emulation calculations 20 times at each energy point, with each run employing a different parameter set randomly generated via Latin Hypercube Sampling (LHS) across the entire parameter space. We then compute the mean of the absolute relative errors at each energy point. The results are clearly visualized in Fig.~\ref{fig:delta_n10Be}, providing a comprehensive assessment of the emulator’s performance across the energy spectrum.

In the upper panel, the phase shifts are shown, with the black solid line representing the exact results and the red dashed line representing the emulation results. Notably, these two curves overlap completely and exhibit a smooth profile, indicating a high degree of reliability in our emulation method. The absence of abrupt changes or anomalies in these curves confirms the robustness of the complex scaling method used in our study.
The lower panel shows the mean of relative errors, which, despite their rapid fluctuations, primarily stabilize around $10^{-7}$. These fluctuations are attributed to the very fine energy grids used in our calculations. Nevertheless, the consistently low error magnitude underscores the emulator's high precision. This level of accuracy, particularly with such small deviations from the exact results, demonstrates the effectiveness of our emulation approach in producing reliable and accurate phase shifts across a broad range of energies.

\section{\label{sec:sum} Summary and conclusions}
In this work, we developed a novel scattering emulator based on the complex scaling method, which significantly enhances the efficiency and accuracy of nuclear reaction analysis. By utilizing a single set of reduced bases, our approach can simultaneously handle multiple channels and potential parameters, leading to substantial reductions in computational storage requirements and faster calculations. We demonstrated the effectiveness of our method through applications to $n$+$^{40}$Ca and $^{11}$Be+$^{64}$Zn elastic scattering, achieving high accuracy and efficiency without encountering the singularity issues common in other methods. The emulator exhibited stable and reliable performance, confirming its robustness and reliability.

In conclusion, our complex scaling-based scattering emulator represents a significant advancement in the field of nuclear reaction research. Its ability to efficiently manage diverse scattering scenarios with high precision and reduced computational demands underscores its potential to become a valuable tool for future studies. This method opens up new possibilities for more comprehensive and detailed investigations in nuclear physics, paving the way for further innovations and discoveries.

\section*{Acknowledgements}
We are grateful to Rimantas Lazauskas for his guidance in implementing the complex scaling method. This work has been supported by the National Natural Science Foundation of China (Grants No. 12105204 and No. 12035011), by the National Key R\&D Program of China (Contract No. 2023YFA1606503), and by the Fundamental Research Funds for the Central Universities.

\bibliography{emulator}

\end{document}